\documentclass[referee]{raa}            

\usepackage{graphicx,times}             
\usepackage{natbib}
\usepackage{amssymb,amsmath}

\bibpunct{(}{)}{;}{a}{}{,}

\begin{document}

   \title{Model Comparison of Dark Energy models Using Deep Network}


   \volnopage{Vol.0 (20xx) No.0, 000--000}      
   \setcounter{page}{1}          

   \author{Shi-Yu Li
      \inst{1}
   \and Yun-Long Li
      \inst{2}
   \and Tong-Jie Zhang
      \inst{1}
   }

   \institute{Department of Astronomy, Beijing Normal University,             Beijing 100875, China; {\it tjzhang@bnu.edu.cn}\\
        \and
             National Space Science Center, Chinese Academy of Sciences,
Beijing 100190, China\\
\vs\no
   {\small Received~~20xx month day; accepted~~20xx~~month day}}

\abstract{ This work uses a combination of a variational auto-encoder and generative adversarial network to compare different dark energy models in light of observations, e.g., the distance modulus from type Ia supernovae. The network finds an analytical variational approximation to the true posterior of the latent parameters in the models, yielding consistent model comparison results with those derived by the standard Bayesian method, which suffers from a computationally expensive integral over the parameters in the product of the likelihood and the prior. The parallel computational nature of the network together with the stochastic gradient descent optimization technique leads to an efficient way to compare the physical models given a set of observations. The converged network also provides interpolation for a dataset, which is useful for data reconstruction.
}

   \authorrunning{S.-Y. Li, Y.-L. Li \& T.-J. Zhang }            
   \titlerunning{Model Comparison of Dark Energy models Using Deep Network}  

   \maketitle

%
%
\section{Introduction}           
\label{sect:intro}
It is well known that predictions about the universe from $\Lambda$CDM are in perfect concordance with observations of the Cosmic Microwave Background (CMB)(\citealt{Aghanim2018}), type Ia supernovae(SNeIa)(\citealt{Betoule2014}), and Baryon Acoustic Oscillations (BAO)(\citealt{Alam2017}), making $\Lambda$CDM the standard paradigm in cosmology. Such a successful model, however, still has its own theoretical problems, which are known as fine tuning and cosmic coincidence(\citealt{Sahni2002, Peebles2003}). More over, a few observations such as the Hubble parameter at high redshift(\citealt{Delubac2015}) and the linear redshift-space distortions(\citealt{Macaulay2013}) have shown tensions with $\Lambda$CDM. All of these motivate research on the universe that allows time-evolving dark energy. People have developed different evolving scalar fields to describe the evolution of dark energy, such as the canonical scalar fields(\citealt{Caldwell1998}) and phantom fields(\citealt{Caldwell2003, Elizalde2004, Scherrer2008}). Various parametrisations of evolving dark energy that broadly describe a large number of scalar field dark energy models are also proposed, such as the Chevallier-Polarski-Linder (CPL) (\citealt{Chevallier2001}) and generalized Chaplygin gas (GCG) models (\citealt{Thakur2012}). Given a specific model and a set of cosmological data, one can study the evolution of the dark energy conveniently.

Then a question of model choice naturally arises with the development of various dark energy models. A variety of methods such as the $F$-test, Akaike information criterion (AIC) (\citealt{Penny2006}), Mallows $C_p$, Bayesian information criterion (BIC) (\citealt{Penny2006}), minimum description length (MDL) (\citealt{Rissanen1978}), and Bayesian model averaging have been proposed to select a good or useful model in light of observations. \citealt{Mackay1992} strongly recommends using Bayesian evidence to assign preferences to alternative models since the evidence is the Bayesian's transportable quantity between models, and the popular easy-to-use AIC and BIC as well as MDL methods are all approximations to the Bayesian evidence (\citealt{Penny2006}). The Bayesian evidence for model selection has been applied to the study of cosmology for a long time (\citealt{Trotta2008, Martin2011, Lonappan2018, Basilakos2018}), and recently a detailed study of Bayesian evidence for a large class of cosmological models taking into account around 21 different dark energy models has been performed by \citealt{Lonappan2018}. Although Bayesian evidence remains the preferred method compared with information criterions, a full Bayesian inference for model selection is very computationally expensive and often suffers from multi-modal posteriors and parameter degeneracies, which lead to a large time consumption to obtain the final result.

The variational auto-encoder (VAE) and the generative adversarial network (GAN) which build upon the variational Bayes theory provide an efficient way to tackle the model selection problem. VAE (\citealt{Kingma2014}) has the ability to approximate the generative process (generate the observed data given the model parameters) and the inference process (infer the model parameters given the observations) which allow one to interpolate between the observed values, thus it is useful in the reconstruction problem. GAN with semi-supervised learning (\citealt{NIPS2014_5423, NIPS2016_6125}) has the ability to effectively learn the distribution of the data, and assign probabilities to different models where the data may come. Thus the combination of VAE and GAN brings us a novel and convenient way to do data reconstruction and model selection at the same time. Since the variational method provides an analytical approximation of the posterior, it is possible to use the fast gradient descent method to find constrains of the parameters rather than using the Monte Carlo Markov Chain approach which may suffer from a low acceptance ratio if the posterior is ill-posed. Moreover, the variational method benefits from natural parallelization of the network computation which can be accelerated by GPU cards.

In this article, we use the VAE-GAN network to learn the distribution of the distance moduli in the $\Lambda$CDM, $\omega$CDM and CPL universe models, then feed the observations of SNeIa to the network to reconstruct dark energy and discriminate the most probable model. The statistical background of the VAE and GAN is briefly reviewed in Section 2, the model structure is described at the end of this section. In Section 3, two toy models are created to test the reconstruction and model discrimination ability of the network. Section 4 describes the observables used in this work, the generation of the training set is introduced. Section 5 reports and discusses the results of the data reconstruction and model comparison given by the network, and some prospects that extend the current work follow the discussion.

\section{The VAE-GAN Network}
\label{sec:method}
The VAE-GAN network proposed by \citealt{pmlr-v48-larsen16} combines a VAE with a GAN, aiming to use the learned feature representations in the GAN discriminator as the basis for the VAE reconstruction, which results in better capturing the data distribution, improving the quality of the inference and the generative process of the network in light of the data. This section briefly reviews the background of the VAE and GAN and then introduces the method to do model selection and data reconstruction using VAE-GAN.

\subsection{The Variational Autoencoder}
\label{subsec:vae}
A VAE (\citealt{Kingma2014}) consists of an encoder $\mathcal{I}$ and a decoder $\mathcal{G}$. The decoder mimics the generative process of a model or a natural phenomenon once given the model parameters or latent variables $\boldsymbol{\xi}$, yielding the likelihood distribution of the data $\tilde{\boldsymbol{x}} \sim \mathcal{G}\left(\boldsymbol{x} \mid  \boldsymbol{\xi}\right)$. The encoder approximates the inverse process that given a set of observations $\boldsymbol{x}$ it infers the posterior distribution of the model parameters or latent variables $\boldsymbol{\xi} \sim \mathcal{I}\left(\boldsymbol{\xi} \mid \boldsymbol{x} \right)$.

The optimal $\mathcal{I}$ and $\mathcal{G}$ are obtained by maximizing the lower bound of the marginal likelihood of the observations via variational bayes (\citealt{Penny2006, Kingma2014}),
\begin{equation}
\mathcal{L}(\boldsymbol{x}) \ge -D_{KL}\left[\mathcal{I}\left(\boldsymbol{\xi} \mid \boldsymbol{x} \right) \Vert p(\boldsymbol{\xi}) \right] + \mathbb{E}_{\mathcal{I}\left(\boldsymbol{\xi} \mid \boldsymbol{x} \right)}\left[ \log \mathcal{G}\left(\boldsymbol{x} \mid  \boldsymbol{\xi}\right) \right]
\label{eq:vae_obj}
\end{equation}
Here, the first item $D_{KL}\left[\cdot\Vert\cdot\right]$ is the Kullback-Leibler (KL) divergence which measures the difference between two distributions. $p(\boldsymbol{\xi})$ is the prior distribution of the latent variables. $\log \mathcal{G}(\boldsymbol{x} \mid \boldsymbol{\xi})$ is the likelihood of the data. The marginal likelihood $\mathcal{L}(\boldsymbol{x})$ equals its lower bound if and only if the approximate posterior $\mathcal{I}(\boldsymbol{\xi} \mid \boldsymbol{x})$ is the same as the true posterior $\mathcal{G}(\boldsymbol{\xi} \mid \boldsymbol{x})$. Eq.\ref{eq:vae_obj} implies that the variational optimal encoder and decoder should constrain the posterior close to the prior while keeping the likelihood as large as possible.

\subsection{The Generative Adversarial Network}
\label{subsec:gan}
A GAN (\citealt{NIPS2014_5423}) consists of a generator $\mathcal{G}$ and a discriminator $\mathcal{D}$. The generator functions similarly to the decoder in that it maps the latent variables $\boldsymbol{\xi} \sim p(\boldsymbol{\xi})$ to the data space $\boldsymbol{x}=\mathcal{G}(\boldsymbol{\xi}) \in p_\mathcal{G}(\boldsymbol{x})$, but the difference is that the mapping is determinant and the sampling process happens only at the latent space. The discriminator assigns probability $\mathcal{D}(\boldsymbol{x})\in\left[0, 1\right]$ to $\boldsymbol{x}$ to tell how probable the real data are (not produced by the generator). The optimal $\mathcal{G}$ and $\mathcal{D}$ are obtained by searching the Nash equilibrium of the minmax game with the value function:
\begin{equation}
\min_\mathcal{G}\max_\mathcal{D}V(\mathcal{G}, \mathcal{D}) =
\mathbb{E}_{p(\boldsymbol{x})}\left[\log\mathcal{D}(\boldsymbol{x}) \right] +
\mathbb{E}_{p_\mathcal{G}(x)}\left[\log(1-\mathcal{D}(\boldsymbol{x})) \right]
\label{eq:gan_obj}
\end{equation}
where $p(x)$ is the distribution of the real data. A small modification of the game (\citealt{NIPS2016_6125}) allows $\mathcal{D}$ to classify $\boldsymbol{x}$ into one of K+1 possible classes, for example, to tell which one of the K classes of dark energy models is the most probable that $\boldsymbol{x}$ is generated from, or $\boldsymbol{x}$ is just the output of the generator which thus belongs to the $(K+1)$-th class.
\begin{equation}
\begin{split}
\min_\mathcal{G}\max_\mathcal{D}\hat{V}(\mathcal{G}, \mathcal{D}) =&
\mathbb{E}_{p(\boldsymbol{x})}\left[\log\mathcal{D}(c \ne K+1 \mid \boldsymbol{x}) \right] +
\mathbb{E}_{p_\mathcal{G}(x)}\left[\log(1-\mathcal{D}(c\ne K+1 \mid \boldsymbol{x})) \right] \\
&+
\mathbb{E}_{p(\boldsymbol{x}, c)}\left[\log \mathcal{D}(c \mid \boldsymbol{x}, c < K+1)\right]
\end{split}
\label{eq:semi_gan_obj}
\end{equation}
Here $c$ is the label of the model. $\mathcal{D}(c \ne K+1 \mid \boldsymbol{x})$ corresponds to $\mathcal{D}(\boldsymbol{x})$ in Eq.\ref{eq:gan_obj}, giving the probability that $\boldsymbol{x}$ is classified as real. $p(\boldsymbol{x}, c)$ is the joint distribution of the real data and the model class. $\mathcal{D}(c \mid \boldsymbol{x}, c < K+1)$ is the probability that $\boldsymbol{x}$ is classified to the right model $c$.

\subsection{Training Algorithm}
\label{subsec:vae-gan}
\begin{figure}
\centering
\includegraphics[width=0.8\textwidth, angle=0]{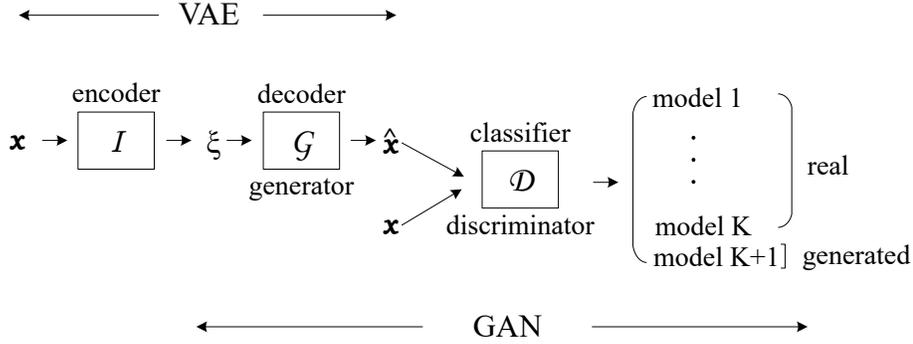}
\caption{The structure of the VAE-GAN network (reproduced from \citealt{pmlr-v48-larsen16} with an additional classifier described in \citealt{NIPS2016_6125}).}
\label{fig:vae_gan}
\end{figure}

The combination of VAE and GAN provides a convenient way to do data interpolation and model selection at the same time, once a set of optimal $\{\mathcal{I}, \mathcal{G}, \mathcal{D}\}$ is obtained by optimizing Eq.\ref{eq:vae_obj} and Eq.\ref{eq:semi_gan_obj}. The basic logic of the VAE-GAN network is shown in Figure \ref{fig:vae_gan}. The observed data $\boldsymbol{x}$ are fed to the encoder $\mathcal{I}$ to find the posterior. Then $\boldsymbol{\xi}$ is sampled from the posterior and fed to the decoder/generator $\mathcal{G}$ to derive the reconstruction (or interpolation) of the input data $\boldsymbol{\hat{x}}$. Finally, the discriminator/classifier extracts the useful features from the reconstruction to derive the probability that $\boldsymbol{\hat{x}}$ belongs to a certain model.

Now the remaining question is that given a set of observations $\{\boldsymbol{x}_i\}_{i=1}^N$ and their covariance $\Sigma_{obs}$ as well as a set of model candidates $\{M_j\}_{j=1}^{K}$, how to find the optimal $\mathcal{I}, \mathcal{G}, \mathcal{D}$ . Since Eq. \ref{eq:vae_obj} and Eq.\ref{eq:semi_gan_obj} set constraints on functions, any flexible functions that have learning abilities can fit in this work. A possible choice is the convolutional neural network (CNN) which is good at representation learning and shift-invariant feature extraction. Suppose $\mathcal{I}, \mathcal{G}, \mathcal{D}$ are CNNs whose parameters are $\boldsymbol{\theta}, \boldsymbol{\phi}, \boldsymbol{\psi}$ respectively. One can generate a batch of training samples from the model candidates and train the networks on these fixed data using stochastic gradient descent
\begin{enumerate}
\item Select $\{\boldsymbol{x}_i, c_i\}$ from the training samples and retrieve the observed part $\boldsymbol{x}_{i, obs}$, $c_j\in\{1, 2, \cdots, K\}$ is the class label. Adding multivariate Gaussian random noise $\boldsymbol{\sigma}_{obs}\sim \mathcal{N}(0, \Sigma_{obs})$ to $\boldsymbol{x}_{i, obs}$ yields $\boldsymbol{x}^*_{i, obs} = \boldsymbol{x}_{i, obs} + \sigma_{obs}$;

\item Feed $\boldsymbol{x}^*_{i, obs}$ to the encoder to get the posterior $\mathcal{I}_{\boldsymbol{\theta}}(\boldsymbol{\xi}\mid\boldsymbol{x}^*_{i, obs})$, then calculate the KL divergence $D_{KL}\left[\mathcal{I}_{\boldsymbol{\theta}}(\boldsymbol{\xi} \mid \boldsymbol{x}^*_{i, obs} ) \Vert p(\boldsymbol{\xi}) \right]$ (corresponding to the first item in Eq.\ref{eq:vae_obj}). Suppose the posterior is a multivariate Gaussian distribution with diagonal covariance, $\mathcal{I}_{\boldsymbol{\theta}}(\boldsymbol{\xi}\mid\boldsymbol{x}^*_{i, obs}) = \mathcal{N}(\boldsymbol{\mu}_i, \mathbf{I}\boldsymbol{\sigma}_i^2)$, and the prior is the standard normal distribution, $p(\boldsymbol{\xi})=\mathcal{N}(\mathbf{0}, \mathbf{I})$, then the KL divergence can be analytically written as $-\frac{1}{2}\sum(\mathbf{1} + \log\boldsymbol{\sigma}_i^2 - \boldsymbol{\mu}_i^2 - \boldsymbol{\sigma}_i^2)$, where the square and sum operations are element-wise (\citealt{Kingma2014}). Find the gradient of the KL divergence with respect to the parameter of the encoder,
\begin{equation}
\Delta \boldsymbol{\theta}_{KL, i} = -\nabla_{\boldsymbol{\theta}}D_{KL}\left[\mathcal{I}_{\boldsymbol{\theta}}(\boldsymbol{\xi} \mid \boldsymbol{x}^*_{i, obs} ) \Vert p(\boldsymbol{\xi}) \right].
\label{eq:update_kl}
\end{equation}

\item Sample $\boldsymbol{\xi}_i$ from the posterior and feed it to the generator to obtain the reconstruction $\boldsymbol{\tilde{x}}_i = \mathcal{G}_{\boldsymbol{\phi}}(\boldsymbol{\xi}_i)$. The observed part of the reconstruction $\tilde{\boldsymbol{x}}_{i, obs}$ together with $\boldsymbol{x}^*_{i, obs}$ and $\Sigma_{obs}$ gives the negative log likelihood $-\log \mathcal{G}_{\boldsymbol{\phi}}(\boldsymbol{x}^*_{i, obs} \mid {\boldsymbol{\xi}_i}) = \frac{1}{2}\chi^2 + const.$ (corresponding to the second item in Eq.\ref{eq:vae_obj}), where $\chi^2 = (\boldsymbol{x}^*_{i, obs} - \tilde{\boldsymbol{x}}_{i, obs})^T \Sigma_{obs}^{-1} (\boldsymbol{x}^*_{i, obs} - \tilde{\boldsymbol{x}}_{i, obs})$ is the goodness of fit that is broadly used in the model regression problems. The $const.$ is the normalization constant of the likelihood, having the value of $\frac{N_{obs}}{2}\log(2\pi) + \frac{1}{2}\log\det\Sigma_{obs}$, where $N_{obs}$ is the dimension of the covariance $\Sigma_{obs}$. Because likelihood depends on both the encoder and the generator, its gradient provides an update to $\mathcal{I}_{\boldsymbol{\theta}},\mathcal{G}_{\boldsymbol{\phi}}$,
\begin{equation}
\begin{split}
\Delta\boldsymbol{\theta}_{\mathcal{L}, i} &= \nabla_{\boldsymbol{\theta}}\log \mathcal{G}_{\boldsymbol{\phi}}(\boldsymbol{x}^*_{i, obs} \mid {\boldsymbol{\xi}_i}),\\
\Delta\boldsymbol{\phi}_{\mathcal{L}, i} &= \nabla_{\boldsymbol{\phi}} \log \mathcal{G}_{\boldsymbol{\phi}}(\boldsymbol{x}^*_{i, obs} \mid {\boldsymbol{\xi}_i}).
\end{split}
\label{eq:update_like}
\end{equation}

\item Sample $\boldsymbol{\xi}_j$ from the prior $p(\boldsymbol{\xi})$ and feed to $\mathcal{G}_{\boldsymbol{\phi}}$ to generate a new sample $\tilde{\boldsymbol{x}}_j$. Feed $\boldsymbol{x}_i, \tilde{\boldsymbol{x}}_i, \boldsymbol{x}_j$ to the discriminator $\mathcal{D}_{\boldsymbol{\psi}}$ to obtain the logits $\boldsymbol{l}_i, \tilde{\boldsymbol{l}}_i, \boldsymbol{l}_j$ which can be interpreted as probabilities, e.g., $\mathcal{D}_{\boldsymbol{\psi}}(k \mid \boldsymbol{x}_i) = \exp (l_{ik}) / \sum_k \exp (l_{ik})$ and $l_{ik}$ is the $k$-th element of the logit $\boldsymbol{l}_i$. Substituting the probabilities into Eq.\ref{eq:semi_gan_obj} yields,
\begin{equation}
\begin{split}
\hat{V}(\mathcal{G}_{\boldsymbol{\phi}}, \mathcal{D}_{\boldsymbol{\psi}}) = \log \mathcal{D}_{\boldsymbol{\psi}}(c = c_i \mid \boldsymbol{x}_i)+
\log\mathcal{D}_{\boldsymbol{\psi}}(c = K+1 \mid \tilde{\boldsymbol{x}}_i) +\log\mathcal{D}_{\boldsymbol{\psi}}(c = K+1 \mid \tilde{\boldsymbol{x}}_j).
\end{split}
\label{eq:semi_gan_mini}
\end{equation}
The gradient of $\hat{V}(\mathcal{G}_{\boldsymbol{\phi}}, \mathcal{D}_{\boldsymbol{\psi}})$ provides a modification to $\mathcal{G}_{\boldsymbol{\phi}}, \mathcal{D}_{\boldsymbol{\psi}}$,
\begin{equation}
\begin{split}
\Delta \boldsymbol{\phi}_{\hat{V}, ij} &= -\nabla_{\phi} \hat{V}(\mathcal{G}_{\boldsymbol{\phi}}, \mathcal{D}_{\boldsymbol{\psi}}), \\
\Delta \boldsymbol{\psi}_{\hat{V}, ij} &= +\nabla_{\psi} \hat{V}(\mathcal{G}_{\boldsymbol{\phi}}, \mathcal{D}_{\boldsymbol{\psi}}).
\end{split}
\label{eq:update_v}
\end{equation}

\item Update the parameters of the encoder, the generator and the discriminator using a learning rate of $\alpha$,
\begin{equation}
\begin{split}
\boldsymbol{\theta} &\leftarrow \boldsymbol{\theta} + \alpha(\Delta \boldsymbol{\theta}_{KL, i} + \Delta \boldsymbol{\theta}_{\mathcal{L}, i}), \\
\boldsymbol{\phi}   &\leftarrow \boldsymbol{\phi} + \alpha(\Delta\boldsymbol{\phi}_{\mathcal{L}, i} + \Delta \boldsymbol{\phi}_{\hat{V}, ij}), \\
\boldsymbol{\psi}   &\leftarrow \boldsymbol{\psi} + \alpha\Delta\boldsymbol{\psi}_{\hat{V}, ij}.
\end{split}
\label{eq:update_params}
\end{equation}
\end{enumerate}

The training process can be easily generalized to mini-batch training to obtain a faster convergence rate. Several training techniques that stabilize or accelerate the training process are also applicable in this problem (\citealt{Radford2015,NIPS2016_6125,Sonderby2016,Mescheder2018ICML}).

The encoder consists of four 1-D convolutional layers and two dense layers. Each layer is followed by a batch renormalization layer (\citealt{ioffe2017}) and an activation layer with Leaky Rectified Linear Unit (a simple variant of ReLU \citealt{Nair2010}), except the last layer which acts as the output. The input of the encoder has a size of 580 which is the number of distance moduli in the Union2.1 dataset, a compilation of SNeIa, later introduced in Section \ref{sec:data}. The dimension of the latent variable should be no less than the number of parameters in the physical models used in the problem, and we set it 20. The size of the convolutional kernel is fixed to 7 and the stride is 4 except for the input layer whose convolutional kernel and stride are of size 69 and 1 respectively. The generator consists of one dense layer and four 1-D fractional convolutional layers. The sizes of the convolutional kernel and stride are the same as the encoder (because it is the inverse process of the encoder), except that the output of the last layer has a dimension of 2048. The discriminator consists of four 1-D convolutional layers and one dense layer. The configuration is similar to the encoder, except that the sizes of the input and output are 2048 and $K+1$ respectively.

\section{Tests On Toy Models}
\label{sec:test}
This section creates two toy models to test the data reconstruction and model comparison ability of the network.

Model 1,
\begin{equation}
\begin{split}
&y = Az^2 + (-A + B)z + C \\
&where, A \sim \mathcal{N}(-4, 0.1), B \sim \mathcal{N}(0, 0.01), C \sim \mathcal{N}(0, 0.1)
\end{split}
\label{eq:model_1}
\end{equation}

Model 2,
\begin{equation}
\begin{split}
&y=A\sin(\omega z) + C \\
&where, A \sim \mathcal{N}(1, 0.1), \omega \sim \mathcal{N}(\pi, 0.01), C \sim \mathcal{N}(0, 0.1)
\end{split}
\label{eq:model_2}
\end{equation}

\begin{figure}
\centering
\includegraphics[width=0.5\textwidth, angle=0]{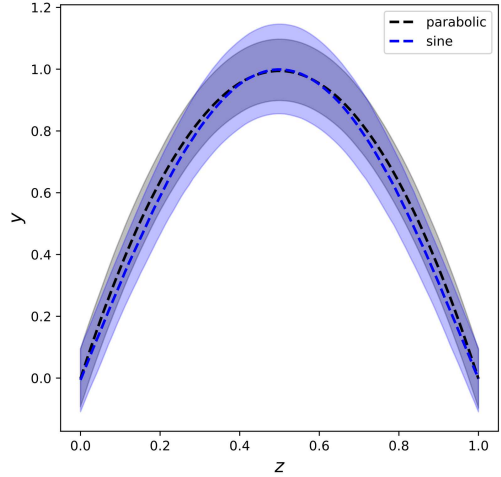}
\caption{The distribution of the outputs of the toy models.}
\label{fig:toymodel}
\end{figure}

Model 1 and Model 2 have similar distributions as shown in Figure \ref{fig:toymodel}. Given the observations $\boldsymbol{x}_{obs, real}$ which are generated by the underlying model $y_{true}=-3.5 z^2 + 3.6 z - 0.1$ on $\boldsymbol{z}_{obs}=\{ z_1, z_2, \cdots, z_{580} \}$ with an error matrix $\Sigma_{obs}$, we would like to fit the two toy models to the observations to tell which one is most probable to be the true model, and interpolate the data with the model at $\boldsymbol{z}^*=\{z_1^*, \cdots, z_M^*\}$, for example, $\boldsymbol{z}^*$ even staying in the interval $[0, 1]$ with $M=1468$.

\begin{figure}
\centering
\includegraphics[width=0.5\textwidth, angle=0]{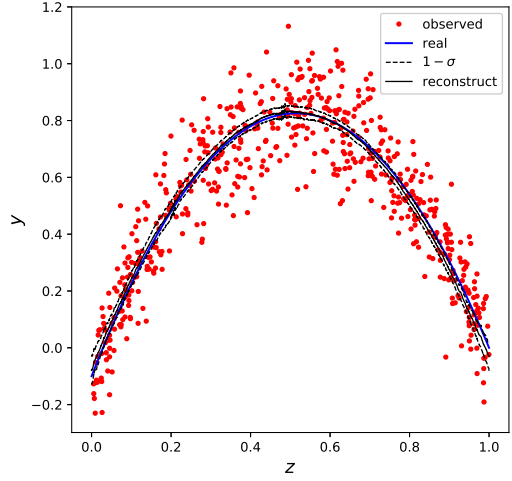}
\caption{Reconstruction of the data. The red dots represent the observed data with noise characterized by $\Sigma_{obs}$. The blue line shows the true model where the observations are generated from. The black line is the reconstruction of the data by the network given the observations.}
\label{fig:toymodel_reconstr}
\end{figure}

First we concatenate and sort $\boldsymbol{z}$ and $\boldsymbol{z}^*$, and call the new one $\boldsymbol{z}$. Then sample $\{A_i, B_i, C_i, \omega_i\}$ from the priors of the toy models and generate the training samples $\boldsymbol{x}_i = M_k(\boldsymbol{z} \mid A_i, B_i, C_i, \omega_i)$ (Note that which set of parameters should be used depends on the toy model). Here 12 800 samples for each model are generated as the training dataset. Finally, the training set $\{\boldsymbol{x}\}_{i=1}^{25600}$ together with the observation error $\Sigma_{obs}$ is fed into the network. Once the training converges, one can put  the observations $\boldsymbol{x}_{obs, real}$ into the network to tell which toy model is most probable and get the interpolation, see Figure \ref{fig:toymodel_reconstr}. In this task, the discriminator has a classification accuracy of almost 1. It assigns a probability of 97\% to the parabolic model (Model 1), which is indeed the case.

\section{The Dataset}
\label{sec:data}
\subsection{The Observations}
The observations are from the Union2.1 compilation (\citealt{Suzuki2012}) which contains 580 SNeIa. Union2.1 provides the distance moduli with their covariance matrix. Let $\boldsymbol{z}_{obs}$ denote the redshift of the 580 SNeIa, and $\boldsymbol{x}_{obs, real}$ signify the measured distance moduli, $\Sigma_{obs}$ represents the covariance of the distance moduli with systematics.

\subsection{The Training Set}
We study the model comparison problem among three dark energy models: (1) $\omega(z)=-1$ ($\Lambda$ CDM); (2) $\omega(z)=\omega_{DE}$ ($\omega$CDM); (3) $\omega(z)=\omega_0 + \omega_a \frac{z}{1+z}$ (CPL), given a set of observations of distance moduli at different redshifts. The expansion rate of a spatially flat FRW universe is determined by the matter and dark energy,
\begin{equation}
H^2(z) = H^2_0\left\{ \Omega_{m0} (1 + z)^3 + (1 - \Omega_{m0})\exp\left[ 3\int\frac{1+\omega(z')}{1+z'}dz' \right] \right\}
\label{eq:hubble}
\end{equation}
The luminosity distance is closely related to the Hubble expansion rate (Eq.\ref{eq:lumdis}), and the distance modulus is given by Eq.\ref{eq:mu}.
\begin{equation}
D_L(z) = c(1+z) \int_0^z dz'\frac{1}{H(z')}
\label{eq:lumdis}
\end{equation}
\begin{equation}
\mu(z) = 5 \log_{10} D_L(z) + 25
\label{eq:mu}
\end{equation}
For each dark energy model, 12 800 samples are generated at the redshift $\boldsymbol{z} = sort\{\boldsymbol{z}_{obs}, \boldsymbol{z}^*\}$, given the priors of the parameters as,
\begin{equation}
\begin{split}
\Omega_{m0} &\sim \mathcal{U}(0.1, 0.9)\\
H_0         &\sim \mathcal{U}(50, 90)\\
\omega_{DE} &\sim \mathcal{U}(-1.8,-0.4)\\
\omega_0    &\sim \mathcal{U}(-1.9, -0.4)\\
\omega_a    &\sim \mathcal{U}(-4.0, 4.0)
\end{split}
\label{eq:prior}
\end{equation}
$\boldsymbol{z}^*$ has 1468 elements evenly located in the interval,  $[0.8\min(\boldsymbol{z}_{obs}), 1.2\max(\boldsymbol{z}_{obs})]$. The $12 800 \times 3$ samples are fixed as the training set.

\section{Results AND Discussions}
\label{sec:result}

\begin{figure}
\centering
\includegraphics[width=0.8\textwidth, angle=0]{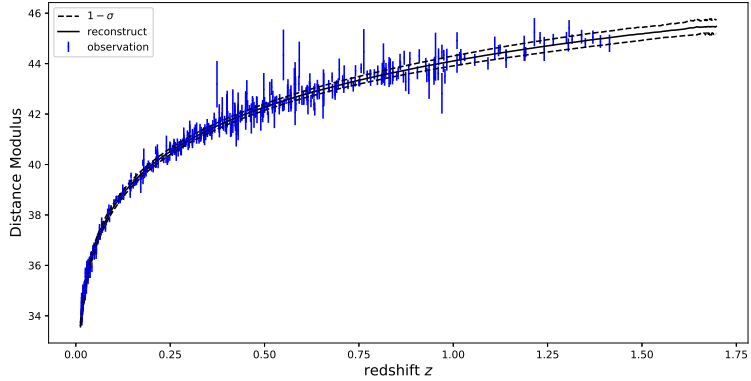}
\caption{Reconstruction of the distance modulus by the network.}
\label{fig:reconstr}
\end{figure}

Figure \ref{fig:reconstr} shows the reconstruction of the distance modulus produced by the network. Given the observed distance modulus $\boldsymbol{\mu}_{obs}$, the discriminator $\mathcal{D}$ assigns probability to each model with,
\begin{equation}
\begin{split}
\mathcal{D}({\Lambda CDM}\mid \boldsymbol{\mu}_{obs}) &= 56.2\%, \\
\mathcal{D}({\omega CDM} \mid \boldsymbol{\mu}_{obs}) &= 28.6\%, \\
\mathcal{D}(CPL \mid \boldsymbol{\mu}_{obs}) &= 15.1\%
\end{split}
\label{eq:result}
\end{equation}
We conclude that $\Lambda$CDM is slightly more favoured than the other two models while CPL is the least favoured in light of the observations. This result is consistent with the one derived by the Bayesian evidence method in {\citealt{Lonappan2018} that finds the log evidence of each model to be $\log\mathcal{Z}_{\Lambda CDM} = -68.11, \log\mathcal{Z}_{\omega CDM} = -69.27$ and $\log\mathcal{Z}_{CPL} = -69.73$. These evidences can be interpreted into probabilities 66.2\%, 20.7\% and 13.1\%, respectively, given the non-informative prior $p(\Lambda CDM)=p(\omega CDM)=p(CPL)$.

The classification accuracy of the three models with the associated observation error is reported as 47.9\%. The accuracy is subjectively low, although it is not unexpected. The integration operations in Eq. \ref{eq:hubble} and Eq. \ref{eq:lumdis} which act as low-pass filters smooth out the local high frequency features that are useful for model comparison. Thus, the convolutional kernel in the network needs to search for useful low frequency features which are less informative once the output distributions of the models overlap each other. This is the case that we meet in this problem. If one uses $p(\boldsymbol{x} \mid M_i)$ to represent the model's prediction of the distribution of the data (it is the evidence of the model), then the theoretical optimal discriminator assigns each model $M_i$ with a probability of $p({\boldsymbol{x} \mid M_i})/\sum_i p({\boldsymbol{x} \mid M_i})$. Once $\boldsymbol{x}_{obs}$ drops in the overlapped region where $p({\boldsymbol{x}_{obs} \mid M_i}) \approx p({\boldsymbol{x}_{obs} \mid M_j}), \forall i, j$, the discriminator loses its ability to discriminate the models confidently.

Note that $\Lambda$CDM is a special case of $\omega$CDM while the latter is a special case of the CPL model, which means there is always a region overlapped among the data distributions of the three models. If the measurements of the distance moduli are accurate enough, the region covered by $p(\boldsymbol{x} \mid \Lambda CDM)$ is negligible compared to $p(\boldsymbol{x} \mid \omega CDM)$, and the latter is negligible compared to $p(\boldsymbol{x} \mid CPL)$. Thus $\boldsymbol{x}$ randomly generated by $\omega CDM$ has an extremely low probability to drop in the region of $p(\boldsymbol{x} \mid \Lambda CDM)$ and conversely a high probability is assigned to $\boldsymbol{x}$ that it comes from $\Lambda CDM$ if it falls in the region of $p(\boldsymbol{x} \mid \Lambda CDM)$. This situation is also applicable to the comparison between $\omega CDM$ and $CPL$. Then the discriminator has a great confidence to tell from which model $\boldsymbol{x}$ comes.

\begin{figure}
\centering
\includegraphics[width=0.8\textwidth, angle=0]{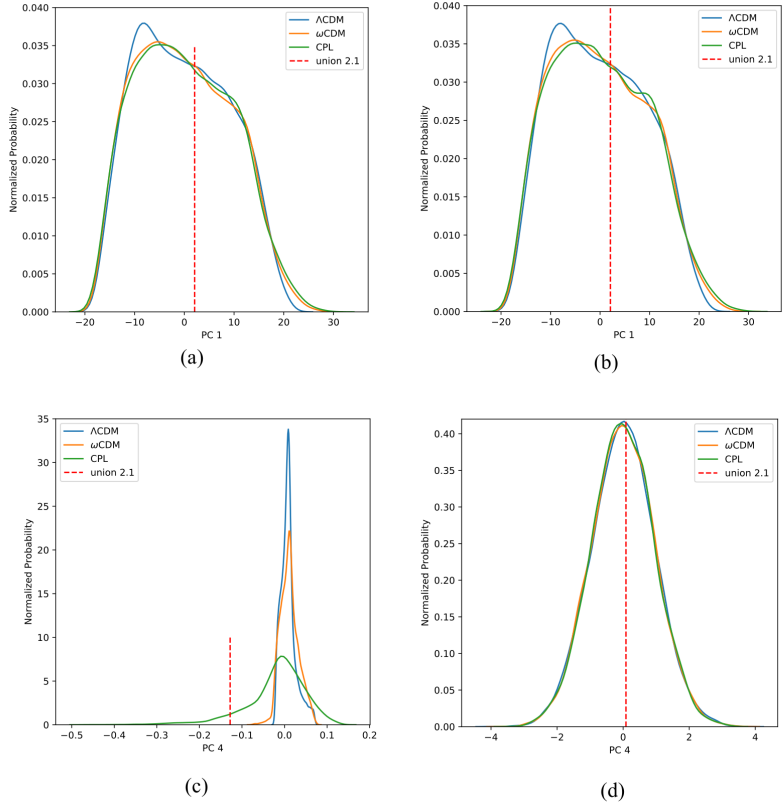}
\caption{Normalized distributions of the projections of training data onto the 1st and 4th PCs. The red dashed line represents the projection of the Union2.1 data to the PCs. (a) The distribution of projection onto the 1st PC with no observation errors; (b) The distribution of the projection onro the 1st PC with the covariance matrix from the Union2.1; (c) The distribution of projection onto the 4th PC with no observation errors; (d) The distribution of the projection onto the 4th PC with the covariance matrix from the Union2.1. (a) and (c) share the same set of PCs, while (b) and (d) share another set of PCs.}
\label{fig:dist_of_pcs}
\end{figure}

Figure \ref{fig:dist_of_pcs} is an illustration of the discussion above.  The left column shows the normalized histograms of projections of the training samples to their 1st and 4th principal components (PCs) with no observation errors. The upper left panel reveals that the low frequency part (1st PC) of the model contributes little to the model discrimination, because the projection of the Union2.1 data to the 1st PC drops in the region where all the models have similar probabilities. The lower left panel demonstrates that the high frequency part (4th PC) of the model is useful for model discrimination, because the projection of the Union2.1 data to the 4th PC is located in the region where the model hasobviously different probabilities.

The discriminator degrades, however, if a non-zero observation error $\Sigma_{obs}$ is involved in the problem. The overlapped region expands due to the errors so that it is not negligible anymore. An extreme limit is that the errors becomes infinity, thus the distribution of the three models become the same so that the discriminator can only make a random guess about which model is true. In this situation the accuracy degrades to $1/3$. Finite observation errors lead to a non-negligible intersection where the discriminator lose the ability to tell confidently from which model the data comes. This is illustrated in the bottom row of Figure \ref{fig:dist_of_pcs}. The lower right panel shows the distribution of 4-th PC scores of the training samples with the covariance matrix of Union2.1. The projection of the Union2.1 data to the 4th PC now locates in the region where the different models have similar probabilities. This explains why the network yields a result that is in good concordance with the standard Bayesian analysis but has a subjectively low classification accuracy - the model classification accuracy is intrinsically determined by the nested structure of the three models as well as the observation noise. The variational network successfully learns the posterior distribution and the likelihood distribution to produce a consistent result.

Although this work uses the distance modulus for model comparison and data reconstruction, it is easy to extend the scenario to Hubble parameters or another dataset. The framework should be further considered to include not only $\boldsymbol{x}_{obs}$ but also its $n$-th derivatives $\boldsymbol{x}^{(n)}_{obs}$, e.g., both the angular diameter distance and the Hubble parameter measured by BAO, to let the encoder and discriminator benefit from different datasets. Another improvement of the framework is to allow the reconstruction of the data to implicitly include a model averaging process which will enhance the generalization of the reconstruction, for example, extend the VAE-GAN to its more powerful variant CVAE-GAN (\citealt{Bao2017}). These are left to future works.

\begin{acknowledgements}
This work was funded by the National Science Foundation of China (Grants No.11573006, 11528306), National Key R\&D Program of China (2017YFA0402600) and the 13th Five-year Informatization Plan of Chinese Academy of Sciences, Grant No. XXH13505-04.
\end{acknowledgements}

\bibliographystyle{raa}
\bibliography{bibtex}
\end{document}